\documentclass[zpreprint,zbstpl]{zeus_paper}
\usepackage[english]{babel}
\newcommand{\ph}        {\phi (1020)}
\def\E{e^+e^-}
\newcommand{\xf}{x_p}
\newcommand{\xb}{X_{\mathrm{B}} }
\newcommand{\yb}{Y_{\mathrm{B}} }
\newcommand{\zb}{Z_{\mathrm{B}} }

\chardef\usc=95
\chardef\til=126

\newcommand{\beq}{\begin{equation}}
\newcommand{\eeq}{\end{equation}  }

\def\citeCTD{{\cite{%
nim:a279:290,*npps:b32:181,*nim:a338:254%
}}\xspace}
\def\citeCAL{{\cite{%
nim:a309:77,*nim:a309:101,*nim:a321:356,*nim:a336:23%
}}\xspace}

\begin{document}

\prepnum{DESY 02-184}

\title{
Observation of the strange sea in the proton via
inclusive  $\phi$-meson production\\
in neutral current deep inelastic scattering\\
at HERA
}

\author{ZEUS Collaboration}
\date{2nd November 2002}

\abstract{
        Inclusive  $\ph$-meson production in neutral
        current deep inelastic $e^+p$ scattering
        has been measured with the ZEUS detector at HERA
        using  an integrated luminosity of 45~pb$^{-1}$.
        The $\phi$ mesons were
        studied in the range $10<Q^2<100$~GeV$^2$, where
        $Q^2$ is the virtuality of the exchanged photon,
        and in restricted kinematic regions in
        the transverse momentum, $p_T$, pseudorapidity, $\eta$,
        and the scaled momentum in the Breit frame, $x_p$.
        Monte Carlo models with the
        strangeness-suppression factor as determined by analyses of
        $\E$ annihilation events overestimate the cross sections.
        A smaller value of the strangeness-suppression factor
        reduces the predicted cross sections, but fails to reproduce the
        shapes of the measured differential cross sections.
        High-momentum $\phi$ mesons in the
        current region of the Breit frame give
        the first direct
        evidence for the strange sea in the proton at low $x$.
}

\makezeustitle

%
%
%
%
\pagenumbering{Roman}                                                                              
                                                   %
\begin{center}                                                                                     
{                      \Large  The ZEUS Collaboration              }                               
\end{center}                                                                                       
  S.~Chekanov,                                                                                     
  D.~Krakauer,                                                                                     
  S.~Magill,                                                                                       
  B.~Musgrave,                                                                                     
  J.~Repond,                                                                                       
  R.~Yoshida\\                                                                                     
 {\it Argonne National Laboratory, Argonne, Illinois 60439-4815}~$^{n}$                            
\par \filbreak                                                                                     
  M.C.K.~Mattingly \\                                                                              
 {\it Andrews University, Berrien Springs, Michigan 49104-0380}                                    
\par \filbreak                                                                                     
  P.~Antonioli,                                                                                    
  G.~Bari,                                                                                         
  M.~Basile,                                                                                       
  L.~Bellagamba,                                                                                   
  D.~Boscherini,                                                                                   
  A.~Bruni,                                                                                        
  G.~Bruni,                                                                                        
  G.~Cara~Romeo,                                                                                   
  L.~Cifarelli,                                                                                    
  F.~Cindolo,                                                                                      
  A.~Contin,                                                                                       
  M.~Corradi,                                                                                      
  S.~De~Pasquale,                                                                                  
  P.~Giusti,                                                                                       
  G.~Iacobucci,                                                                                    
  A.~Margotti,                                                                                     
  R.~Nania,                                                                                        
  F.~Palmonari,                                                                                    
  A.~Pesci,                                                                                        
  G.~Sartorelli,                                                                                   
  A.~Zichichi  \\                                                                                  
  {\it University and INFN Bologna, Bologna, Italy}~$^{e}$
% ...
\newpage                                                                                           
$^{\    1}$ on leave of absence at University of                                                   
Erlangen-N\"urnberg, Germany\\                                                                     
$^{\    2}$ supported by the GIF, contract I-523-13.7/97 \\                                        
% ...
                                                           %
                                                           %
\newpage   
                                                           %
                                                           %
\begin{tabular}[h]{rp{14cm}}                                                                       
$^{a}$ &  supported by the Natural Sciences and Engineering Research                               
          Council of Canada (NSERC) \\                                                             
$^{b}$ &  supported by the German Federal Ministry for Education and                               
          Research (BMBF), under contract numbers HZ1GUA 2, HZ1GUB 0, HZ1PDA 5, HZ1VFA 5\\         
$^{c}$ &  supported by the MINERVA Gesellschaft f\"ur Forschung GmbH, the                          
          Israel Science Foundation, the U.S.-Israel Binational Science                            
          Foundation and the Benozyio Center                                                       
          for High Energy Physics\\                                                                
$^{d}$ &  supported by the German-Israeli Foundation and the Israel Science                        
          Foundation\\                                                                             
$^{e}$ &  supported by the Italian National Institute for Nuclear Physics (INFN) \\                
$^{f}$ &  supported by the Japanese Ministry of Education, Science and                             
          Culture (the Monbusho) and its grants for Scientific Research\\                          
$^{g}$ &  supported by the Korean Ministry of Education and Korea Science                          
          and Engineering Foundation\\                                                             
$^{h}$ &  supported by the Netherlands Foundation for Research on Matter (FOM)\\                   
$^{i}$ &  supported by the Polish State Committee for Scientific Research,                         
          grant no. 620/E-77/SPUB-M/DESY/P-03/DZ 247/2000-2002\\                                   
$^{j}$ &  partially supported by the German Federal Ministry for Education                         
          and Research (BMBF)\\                                                                    
$^{k}$ &  supported by the Fund for Fundamental Research of Russian Ministry                       
          for Science and Edu\-cation and by the German Federal Ministry for                       
          Education and Research (BMBF)\\                                                          
$^{l}$ &  supported by the Spanish Ministry of Education and Science                               
          through funds provided by CICYT\\                                                        
$^{m}$ &  supported by the Particle Physics and Astronomy Research Council, UK\\                   
$^{n}$ &  supported by the US Department of Energy\\                                               
$^{o}$ &  supported by the US National Science Foundation\\                                        
$^{p}$ &  supported by the Polish State Committee for Scientific Research,                         
          grant no. 112/E-356/SPUB-M/DESY/P-03/DZ 301/2000-2002, 2 P03B 13922\\                    
$^{q}$ &  supported by the Polish State Committee for Scientific Research,                         
          grant no. 115/E-343/SPUB-M/DESY/P-03/DZ 121/2001-2002, 2 P03B 07022\\                    
\end{tabular}                                                                                      
                                                           %
                                                           %

\pagenumbering{arabic} 
\pagestyle{plain}
\section{Introduction}
\label{sec-int}

The total quark content of the proton 
has been well determined
\cite{epj:c12:375,*hep-ph-0201195,epj:c4:463,epj:c5:461,epj:c21:33,*zeus_pdf}
through analyses of inclusive deep inelastic scattering (DIS) data.
However,
the flavour decomposition of the sea  
is less well known. 
So far, experimental  constraints
on the strange-quark content of the nucleon
have come from fixed-target neutrino experiments
\cite{zfp:c65:189,*pr:d64:112006,*proc:dis00:93}, which
indicate that the $s\bar{s}$ is suppressed with respect
to the $u\bar{u}$ and $d\bar{d}$ sea by a factor of about two.   
This paper reports a study of the production of
$\phi$-mesons in neutral current $e^+p$
DIS and explores its sensitivity
to the strange sea of the proton at low $x$. 

\begin{figure}
\begin{center}

\vspace{-1.3cm}
\includegraphics[height=11.0cm]{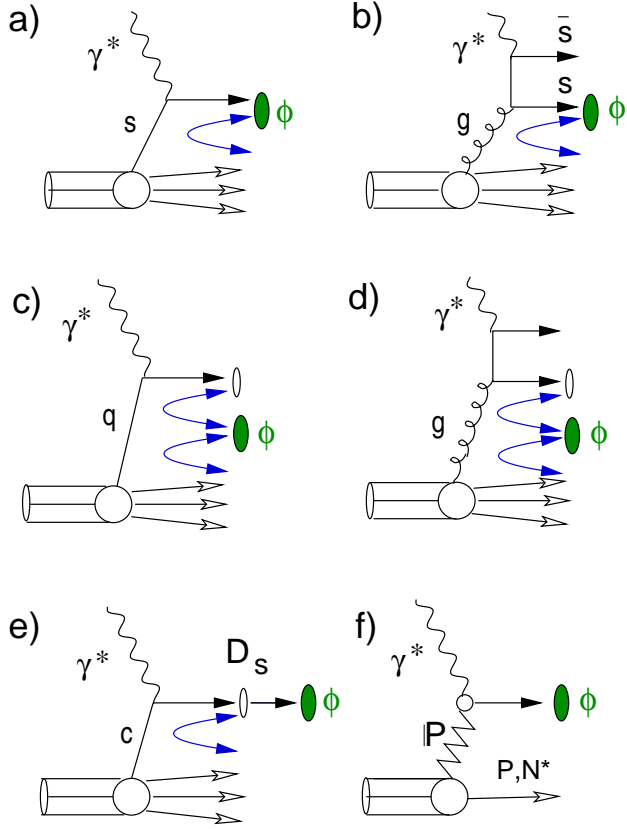}%
\caption{
A schematic representation of different  
mechanisms for $\phi$ production in inclusive DIS:
a) a $\phi$ meson is produced from a strange
quark after the interaction on the strange sea according to the QPM;
b)  a $\phi$ meson is produced from a strange quark emerging from
the BGF process;
c)-d) a $\phi$ meson is produced solely by the hadronisation process,
independent of the flavour of the quark  participating in the hard interaction.
Additional sources  for $\phi$ mesons are the hadronisation
of strange quarks produced by
higher-order gluon splittings, resonance decays, 
such as e) the $D_s$-meson decays; and f) diffractive
$\phi$-meson production. }  
\label{fey}
\end{center}
\end{figure}

Several mechanisms lead to $\phi$-meson production in DIS. 
The $\phi$ meson, which is a nearly pure $s\bar{s}$ state,
can be produced by the 
hadronisation of a strange quark created in the hard scattering 
process of a virtual photon on the strange
sea of the proton, $\gamma^* s \rightarrow s$, as 
illustrated in Fig.~\ref{fey}a). 
The underlying hard-scattering process is 
either zeroth order in QCD, 
namely the quark-parton model 
(QPM), or  first order, 
$\gamma^* s \rightarrow sg$, the QCD
Compton reaction (QCDC).  
Another source of strange quarks is 
boson-gluon fusion (BGF),  
$\gamma^* g \rightarrow s \overline{s}$, Fig.~\ref{fey}b).
In contrast to the QPM and QCDC processes, 
the rate of BGF events is related to the density of gluons in the proton 
and is, therefore, not directly dependent on 
the intrinsic sea-quark content of the proton.
The hadronisation process alone, 
without strange quarks being involved in the hard scattering, 
contributes to the production of $\phi$ mesons, 
as shown in Figs.~\ref{fey}c)-d).
In this case,  $\phi$ mesons are formed from strange quarks 
created during hadronisation. 
Hadronisation of strange quarks produced in 
higher-order QCD reactions related to the splitting
of gluons, $g \rightarrow s \overline{s}$, 
and the decay of higher-mass states, 
such as the $D_s$ meson (Fig.~\ref{fey}e)), 
also contribute.
In addition, diffractive scattering can produce $\phi$ mesons in
the final state (Fig.~\ref{fey}f)). 

Strange-particle production in inclusive DIS has been 
studied at HERA using $K^0$ mesons
and $\Lambda$ baryons \cite{zfp:c68:29,np:b480:3}. 
However, their production rates are dominated by  
the fragmentation process and by the decays of high-mass states, and    
are, therefore, insensitive to the
presence of strange quarks in the hard scattering process. 
For $\phi$ mesons, the contribution from 
resonance decays is relatively small.  
Furthermore, selecting $\phi$ mesons with large longitudinal 
momenta in the Breit frame \cite{feynman:1972:photon, *zpf:c2:237}  
enhances the contribution from the QPM process of Fig.~\ref{fey}a).

In this study, the $\phi$ mesons were identified through the
decay $\phi \rightarrow K^+K^-$.
Their differential cross sections 
are presented as functions of $Q^2= -q^2 = -(k-k')^2$ and
Bjorken $x=Q^2/(2P ·q)$, where  
$k$ and $k'$ are the four-momenta of the initial and 
scattered lepton and $P$ is the
four-momentum of the incoming proton, as well as other variables
that characterise the $\phi$-meson  production.  

\section{Properties of $\phi$ mesons in  the Breit frame}
\label{sec:intr1}

The Breit
frame \cite{feynman:1972:photon,*zpf:c2:237}  
provides a natural system to separate
the radiation of the outgoing
struck quark  from the proton remnant. In this frame,
the exchanged virtual boson with virtuality $Q$ is 
space-like and has a momentum $q=(q_0,q_{\xb},q_{\yb},q_{\zb} )=(0,0,0,-Q)$.
In the QPM, the incident quark has $p_{\zb}=Q/2$
and the outgoing struck quark carries $p_{\zb}=-Q/2$.
All particles with negative $p_{\zb}$
form the current region. These  particles
are produced by the fragmentation   of
the struck quark, so that this region is
analogous to a
single hemisphere of  an $\E$ annihilation event.

The  $\phi$-meson cross sections are  presented
as a function 
of the scaled momentum, $\xf =2\> p /Q$,
where $p$ is the absolute momentum of
the $\phi$ meson in the Breit frame.
In the QPM process, $\gamma^*s \to s$, 
this variable is equal to unity for the $s$-quarks in
the current region. 
As a consequence, leading $\phi$ mesons in the current region with
$\xf$ values close to unity
are a measure of  the hard scattering
of a virtual photon on the strange sea. 
Gluon radiation and the fragmentation process
generally lead to particles with $\xf < 1$, 
and, much less frequently, to $\xf > 1$.

In the target region, 
$\xf$ can be significantly larger than unity. This is because
the maximum momentum of the proton remnant in the QPM
is $Q(1-x)/2x$, therefore $x_p^{\mathrm{max}} = (1-x)/x$.
The $\phi$ mesons in the target region are mostly produced by
the hadronisation processes of Fig.1c)-d), as well as the   
hadronisation of strange quarks from the BGF diagram of Fig.~1b).

\section{Data sample and analysis procedure}

\subsection{Experimental setup}

During the 1995-1997 period,
$45.0\pm 0.7$  pb$^{-1}$ of data were taken 
with the ZEUS detector with
a positron beam energy of $27.5$ GeV and a proton beam
energy of $820$ GeV.

ZEUS is a multipurpose detector described in detail
elsewhere \cite{zeus:1993:bluebook}.
Of particular importance in the
present study  are the central tracking detector
and the calorimeter.

The central tracking detector (CTD) \citeCTD is  a cylindrical
drift chamber with nine superlayers covering the  
polar-angle\footnote{
The ZEUS coordinate system is a right-handed Cartesian system, with
the $Z$ axis pointing in the proton beam direction, referred to
as the ``forward direction'', and the $X$ axis
pointing left towards the centre of HERA. The coordinate origin is at the
nominal interaction point. The pseudorapidity is defined as 
$\eta = -\ln ( \tan \frac{\theta}{2} ) $, where the polar angle, $\theta$, is measured with
respect to the proton beam direction.}
region $15^o < \theta < 164^o$ and the
radial range $18.2 - 79.4$ cm. Each superlayer consists
of eight sense-wire layers. The transverse-momentum resolution
for charged tracks traversing all CTD layers is
$\sigma(p_{T})/p_{T} = 0.0058 p_{T}
\oplus 0.0065 \oplus  0.0014/p_{T}$, 
with $p_{T}$ in GeV.

The CTD
is  surrounded by the uranium-scintillator 
calorimeter, CAL \citeCAL, which is divided
into three parts: forward, barrel and rear.
The calorimeter is longitudinally segmented into electromagnetic
and hadronic sections. 
The smallest subdivision of the CAL is called a
cell. The energy resolution of the calorimeter 
under test-beam conditions
is $\sigma_E/E=0.18/\sqrt{E}$ for electrons and
$\sigma_E/E=0.35/\sqrt{E}$ for hadrons (with $E$ in GeV).

The position of positrons scattered at small angles to the positron beam
direction was measured using the small-angle rear tracking detector 
(SRTD) \cite{nim:a401:63, epj:c21:443}.  
The energy of the scattered positrons was corrected
for the energy loss in the material between the interaction
point and the calorimeter using a presampler (PRES) \cite{nim:a382:419, epj:c21:443}.

\subsection{Kinematic reconstruction and event selection}
\label{sub:ds}

The scattered-positron candidate was identified from the pattern  of
energy deposits in the CAL \cite{nim:a365:508}.   
The kinematic variables, $Q^2$ and $x$, 
were reconstructed by  the following methods:

\begin{itemize}
\item
the electron method (this method is denoted by  the subscript $e$) uses
measurements of the energy and angle
of the scattered positron;

\item
the double angle (DA) method \cite{proc:hera:1991:23,*h1_da} relies on
the angles of the scattered  positron  and of the hadronic energy flow;

\item
the Jacquet-Blondel (JB)  method \cite{proc:epfacility:1979:391} is based entirely on
measurements of the hadronic system.
\end{itemize}
The DIS event selection
was based on the following requirements:

\begin{itemize}

\item[$\bullet$]
$E_{e^{'}}\geq 10$ GeV, where $E_{e^{'}}$ is the energy
of the scattered positron in the calorimeter after the 
correction by the PRES;

\item[$\bullet$]
$10 <  Q^2_e < 100$ GeV$^2$. The upper cut on $Q^2_e$ 
was used  
to reduce the combinatorial background in the $\phi$-meson reconstruction;

\item[$\bullet$]
$40 <  \delta <  60$ GeV, where
$\delta=\sum E_i(1-\cos\theta_i)$,   
$E_i$ is the energy of the $i$th calorimeter 
cell, $\theta_i$ is its angle, and the sum runs over all cells.
This cut further reduces the background from photoproduction and
events with large initial-state radiation;

\item[$\bullet$]
$y_{e}\> \leq\> 0.95$, to remove
events with fake  scattered positrons; 
 
\item[$\bullet$]
$y_{\mathrm{JB}}\> \geq\>  0.04$, to improve
the accuracy of the DA reconstruction used in systematic checks;

\item[$\bullet$]
a primary vertex position, determined from the
tracks fitted to the vertex, in the range
$\mid Z_{\mathrm{vertex}} \mid  < 50$ cm, to reduce
background events from non-$ep$ interactions;

\item[$\bullet$]
the impact point ($X$, $Y$) of the scattered positron in the calorimeter  
must be within a radius $\sqrt{X^2+Y^2}>25$ cm. 

\end{itemize}

The reconstruction of the Breit frame and  
the $Q^2$ and $x$ variables 
was performed using the electron method, since it has the best
resolution at the relatively low $Q^2$ values of this data set.  

\section{Selection of $\phi$ candidates}

Charged tracks measured by the CTD and assigned
to the primary event vertex were selected.
Tracks were required to pass through at least three 
CTD superlayers and have 
transverse momenta $p_{T}>200$  MeV in the laboratory frame, thus   
restricting  
the study to  a CTD region where track acceptance and resolution are high.  

All pairs of oppositely charged tracks were combined   
to form the $\phi$ candidates. 
The  tracks were  assigned the mass of a charged kaon  
when calculating the invariant mass, $M(K^+K^-)$, of each track pair.
The events with $\phi$-meson candidates were
selected using the following requirements:

\begin{enumerate}

\item[$\bullet$]
$0.99 < M( K^+K^- )<1.06 $ GeV;

\item[$\bullet$]
$p_T^{\phi} >1.7$ GeV and $-1.7 < \eta^{\phi}< 1.6$,   
where $p_T^{\phi}$  
is the transverse momentum and $\eta^{\phi}$ is the  
pseudorapidity
of the $\phi$ meson in the laboratory frame.   

\end{enumerate}

The asymmetric cut on $\eta^{\phi}$ was used to avoid
the very forward region that has large track multiplicities,  
resulting in high combinatorial backgrounds. 

\begin{figure}
\begin{center}
\includegraphics[height=12.0cm]{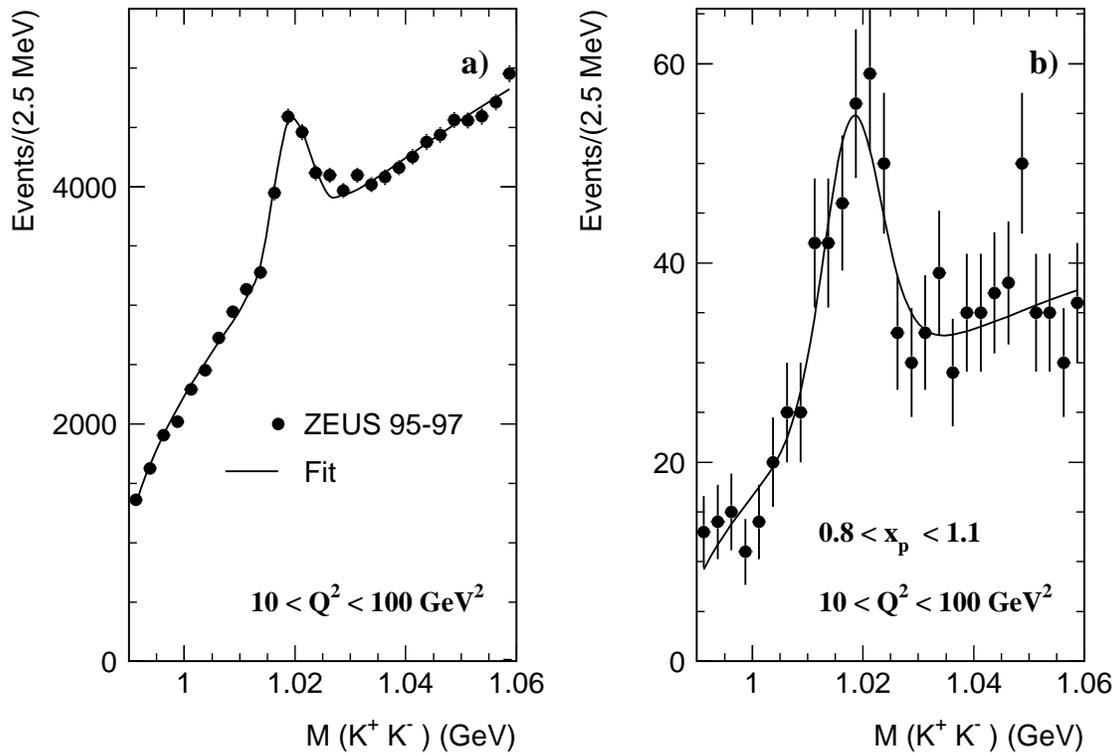}%
\caption{
The invariant mass of the $\phi$-meson 
candidates (points with statistical error bars) 
a) in the restricted kinematic regions 
$p_T^{\phi}>1.7$~GeV and $-1.7<\eta^{\phi}<1.6$;
b) for  the
highest $x_p$ value  in the current region of the Breit frame, in 
addition to the cuts as in  a).  
The solid
lines show the results of the fit described in the text.
}
\label{fig2}
\end{center}
\end{figure}

Figure~\ref{fig2}a) shows the invariant-mass  distribution  
for $\phi$ candidates in the range  $10<Q^2<100$ GeV$^2$.
The invariant mass for the leading $\phi$ mesons 
in the current region of the Breit frame, $0.8<\xf <1.1$, is 
presented in Fig.~\ref{fig2}b). For the latter case,
the DIS events containing the $\phi$-meson candidates have $x<0.006$.  
The solid line in each figure is a fit   
using a relativistic Breit-Wigner (BW) function convoluted
with a Gaussian distribution plus    
a term describing the background: 
$$
F(M) = (BW)\otimes(Gaussian)   
+ 
a·(M - 2 m_K )^{b},  
\label{fit} 
$$
where $a$ and $b$  are free parameters and $m_K$ is the kaon mass.
The fit function contains five free parameters:
normalisation, peak position, width of the Gaussian distribution,
and two parameters describing the background.
When the peak position was left free,
the resulting fit gave
$1019.2\pm 0.3$~MeV,
in agreement with the
PDG value of $1019.456\pm 0.020$~MeV \cite{pdg_2002}.
The width of the Gaussian was $1.6\pm 0.3$ MeV, consistent
with the tracking resolution.
In order to improve the stability of the fit for  
the calculations of the differential cross sections,
the mass peak and width of the Breit-Wigner function were  fixed at the
PDG values \cite{pdg_2002}.
The total number of $\phi$-meson  candidates  determined
from this fit was $4950\pm 214$, while the number of $\phi$-meson candidates 
for the high $\xf$ region was $181\pm 28$.

\section{Event simulations} 
\label{sec:evsim}

A good understanding of hadronisation is a pre-requisite
for the interpretation of the measured
inclusive $\phi$-meson cross sections.
At present, only Monte Carlo (MC) models based on leading-order QCD 
are available to compare
with the experimental results, so that the predictions for the rates of 
$s\bar{s}$ production are plagued by
large model-dependent  uncertainties. 
In MC  models based on the Lund string
fragmentation \cite{prep:97:31},  the production ratio of strange to
light non-strange quarks is parameterised by the
strangeness-suppression factor,
$\lambda_s = P_s/P_{u,d}$, where
$P_s$ ($P_{u,d}$) is the probability of creating $s$ ($u,d$)
quarks in the colour field during fragmentation.
The processes shown in Figs.~\ref{fey}a) and b) are proportional
to $\lambda_s$, while the contributions illustrated in
Figs.~\ref{fey}c) and d) are  
proportional to $\lambda_s^2$.

In $\E$ annihilation, the production of $\phi$-mesons has been well
described using  $\lambda_s = 0.3$ \cite{zpf:c56:521, 
*zfp:c68:1, *zfp:c69:379, *zfp:c73:61}.
However, there are new indications that a larger value, 
$\lambda_s \simeq  0.4$,  may be  
needed \cite{epj:c16:407}, or even 
that a single value cannot accommodate all of the
SLD strangeness-production data \cite{pr:d59:052001}.
When using the same hadronisation model in $e^+p$ scattering,
the measured $K^0$ and 
$\Lambda$ production rates in DIS \cite{zfp:c68:29, np:b480:3} 
and photoproduction  \cite{epj:c2:77} indicate
the need for a smaller value, $\lambda_s\simeq 0.2$.

The measured cross sections 
were compared to various leading-order  MC models 
based  on the QCD parton-cascade approach,  to incorporate
higher-order QCD effects,  followed by  
fragmentation into hadrons. 
The MC events were  generated with
LEPTO 6.5 \cite{cpc:101:108},
ARIADNE 4.07  \cite{cpc:71:15} and HERWIG 6.2 \cite{cpc:67:465} using the
default parameters in each case.
The fragmentation in LEPTO and ARIADNE
is  simulated using the Lund string model \cite{prep:97:31}
as implemented in PYTHIA  \cite{cpc:82:74}, 
whereas the hadronisation stage
in HERWIG is described by a cluster 
fragmentation model \cite{np:b238:492,*np:b310:461}.

The acceptance was calculated using 
ARIADNE, which was interfaced with HERACLES 4.5.2 \cite{cpc:69:155} 
using the DJANGOH program \cite{spi:www:djangoh11}
in order to incorporate first-order electroweak corrections.
The generated events were then  passed  through a full 
simulation of the detector using GEANT 3.13 \cite{tech:cern-dd-ee-84-1} 
and processed with the same
reconstruction program as used for the data.
The detector-level MC samples  were then  
selected in the same way as the data. 

The natural width of the Breit-Wigner distribution
for $\phi$-meson decays was 
set to the default value \cite{pdg_2002} in LEPTO and ARIADNE. 
The HERWIG model sets the particle-decay width 
to zero 
and is therefore less realistic
for the acceptance calculations. The HERWIG model was 
used only for comparisons with the final cross sections. 

The inclusive $\phi$-meson sample contains
a contribution from diffractive processes, which is not well   
simulated in the MC models mentioned above. 
These processes are  characterised by a rapidity gap, chosen 
as $\eta_{\mathrm{max}}<2$,
where $\eta_{\mathrm{max}}$ is defined
as the pseudorapidity
of the energy deposit in the CAL above $400$ MeV closest to the
proton direction, and  by
the presence in the CTD of only a few tracks. 
Diffractive events with $\phi$ mesons were 
generated with PYTHIA 5.7 \cite{cpc:82:74} and  
passed  through
the same simulation of the detector as for inclusive MC events. 
The MC distributions were fit to the data by varying the fraction
of the diffractive $\phi$-meson events from PYTHIA 
and minimising the $\chi^2$   
to obtain  good agreement for
the multiplicity of charged tracks in the CTD.     
The fraction of PYTHIA events needed to obtain good agreement between data
and MC  was  $2.7\pm 0.2\%$ of  the total number of reconstructed 
$\phi$-meson events. 
It was verified that 
this fraction gives a satisfactory description
of the $\phi$-meson events for $\eta_{\mathrm{max}}<2$.  
          
\section{Definition of cross sections and systematic uncertainties }

The $\phi$-meson cross sections were measured  
in the kinematic region  $10<Q^2<100$ GeV$^2$,
$2\cdot 10^{-4} < x < 10^{-2}$,
$1.7 < p_T^{\phi} <7$ GeV and $-1.7 < \eta^{\phi } < 1.6$. 
The cross sections as a function of a given observable, $Y$, were
determined  from   
\begin{equation}
\frac {d\sigma}{dY} = \frac {N } {A \cdot \mathcal {L} \cdot B \cdot 
\Delta Y} \>\> ,
\nonumber
\end{equation}
where $N$ is the number of events with a $\phi$-meson candidate
in a bin of size $\Delta Y$, 
$A$ is the correction factor (which takes into
account migrations, efficiencies and radiative effects for that bin)
and $\mathcal {L}$ is the integrated luminosity.
The  branching ratio, $B$,
for the decay channel $\phi \to K^+K^-$
was taken to be $0.492^{+0.006}_{-0.007}$ \cite{pdg_2002}. 

The acceptance for each kinematic bin was calculated as
$\mathcal{A}^{\mathrm{rec}} / \mathcal{A}^{\mathrm{gen}}$,
where $\mathcal{A}^{\mathrm{rec}}$ ($\mathcal{A}^{\mathrm{gen}}$)  
is the reconstructed (generated) number of events
with $\phi$ mesons. For the calculation of the acceptance,
$2.7\%$ of the total number of inclusive DIS events generated with ARIADNE 
were replaced by diffractive events from PYTHIA.
While the contribution from diffractive $\phi$-meson events is 
negligible for the full phase-space region,
it is important for the high $x_p$ region in the Breit frame, since 
$72\%$ of the
diffractive $\phi$-meson events have  $x_p >0.8$.

The systematic uncertainties of the measured cross sections  
were estimated from the following (the typical contribution
from each item to the uncertainty of the total cross section
is indicated in parentheses): 

\vspace{0.2cm}
 
\begin{itemize}

\item[$\bullet$]
event reconstruction and selection. Systematic
checks were performed by changing the cuts on
$y_{e}$, $y_{\mathrm{JB}}$, $\delta$
and the vertex-position requirement: 
$y_{e}\> \leq\>  0.90$ ($-0.1\%$), 
$y_{\mathrm{JB}}>0.05$ ($-0.05\%$),  
$42 \>  \leq \> \delta\>  \leq\>  58$ GeV ($-0.3\%$),  
$\mid Z_{\mathrm{vertex}} \mid  < 45$ cm ($+0.4\%$).
The radius cut for the position of the scattered positron in the calorimeter  
was raised by 1 cm ($-0.5\%$). 
The minimum accepted 
energy of the scattered positron was  increased  by 1 GeV ($-0.1\%$).
The positron energy  scale
was changed within its $\pm 2\%$ uncertainty ($^{+0.1}_{+0.7}\> \%$); 

\item[$\bullet$]
the DA method was used to reconstruct the Breit frame ($+0.3\%$) 
and the kinematic variables ($+0.08\%$); 

\item[$\bullet$]
the minimum transverse momentum for  $K$-meson candidates 
was raised  by 100 MeV ($+0.6\%$). 
Tracks were required to have  $\mid \eta \mid < 1.75$, in
addition to the requirement of three CTD superlayers ($+0.02\%$); 

\item[$\bullet$]
the form of the background in the fits was changed to a second-order polynomial
function ($+0.4\%$);

\item[$\bullet$]
the fraction of  diffractive $\phi$-meson events in the Monte Carlo sample  
was varied in the range $1.9-3.5\%$ ($\pm 0.03\%$).  

\end{itemize}

\vspace{0.2cm}

The overall systematic uncertainty for the differential cross sections 
was  determined by adding
the  above  uncertainties in quadrature.
The normalisation uncertainty 
due to that of the luminosity measurement, which is $1.6\%$,   
was only added to the overall
systematic uncertainty for the total $\phi$-meson cross section.
The uncertainty in  the $\phi\to K^+K^-$ decay branching ratio
was not included. 

\section{Results}

The overall $\phi$-meson acceptance  
for $10<Q^2<100$ GeV$^2$,  
$2\cdot 10^{-4} < x < 10^{-2}$,  
$1.7 < p_T^{\phi} <7$ GeV and $-1.7 < \eta^{\phi } < 1.6$,  
estimated with DJANGOH, was 45\%. 
The total $\phi$-meson cross section in this  region  is  
$$
\sigma(e^+p \rightarrow e^+\phi X) =
0.507 \pm 0.022(\mbox{stat.})^{+0.010}_{-0.008}(\mbox{syst.}) 
\mbox{ nb}. 
\label{crrr}
$$
This cross section is lower than that predicted by the LEPTO (0.680~nb) 
and ARIADNE (0.701~nb)  
models with the CTEQ5D structure function and with the LEP default  
value of the strangeness-suppression factor, $\lambda_s=0.3$. 
The HERWIG 6.2 model for neutral current DIS processes 
underestimates the measured $\phi$-meson cross section,
predicting 0.36~nb.

In previous studies of neutral kaons and $\Lambda$ baryons
at HERA~\cite{zfp:c68:29, np:b480:3,epj:c2:77}, it
was found that decreasing $\lambda_s$ from its
standard value 
improved  the agreement between the Lund  MC models and the data.
A smaller value of the strangeness-suppression 
factor, $\lambda_s=0.22$,
resulted in an inclusive $\phi$-meson cross section
of 0.501(0.509)~nb for LEPTO (ARIADNE),
which agrees well with the present measurement. 
Therefore, $\lambda_s=0.22$   
was  used as the default
for LEPTO and ARIADNE in the following comparisons. 
A comparison of the data with the predicted 
cross sections gave an uncertainty of $\pm 0.02$ on the
$\lambda_s$ value used in this analysis.  

\subsection{Differential $\phi$-meson cross  sections}

\begin{figure}
\begin{center}
\includegraphics[height=19.0cm]{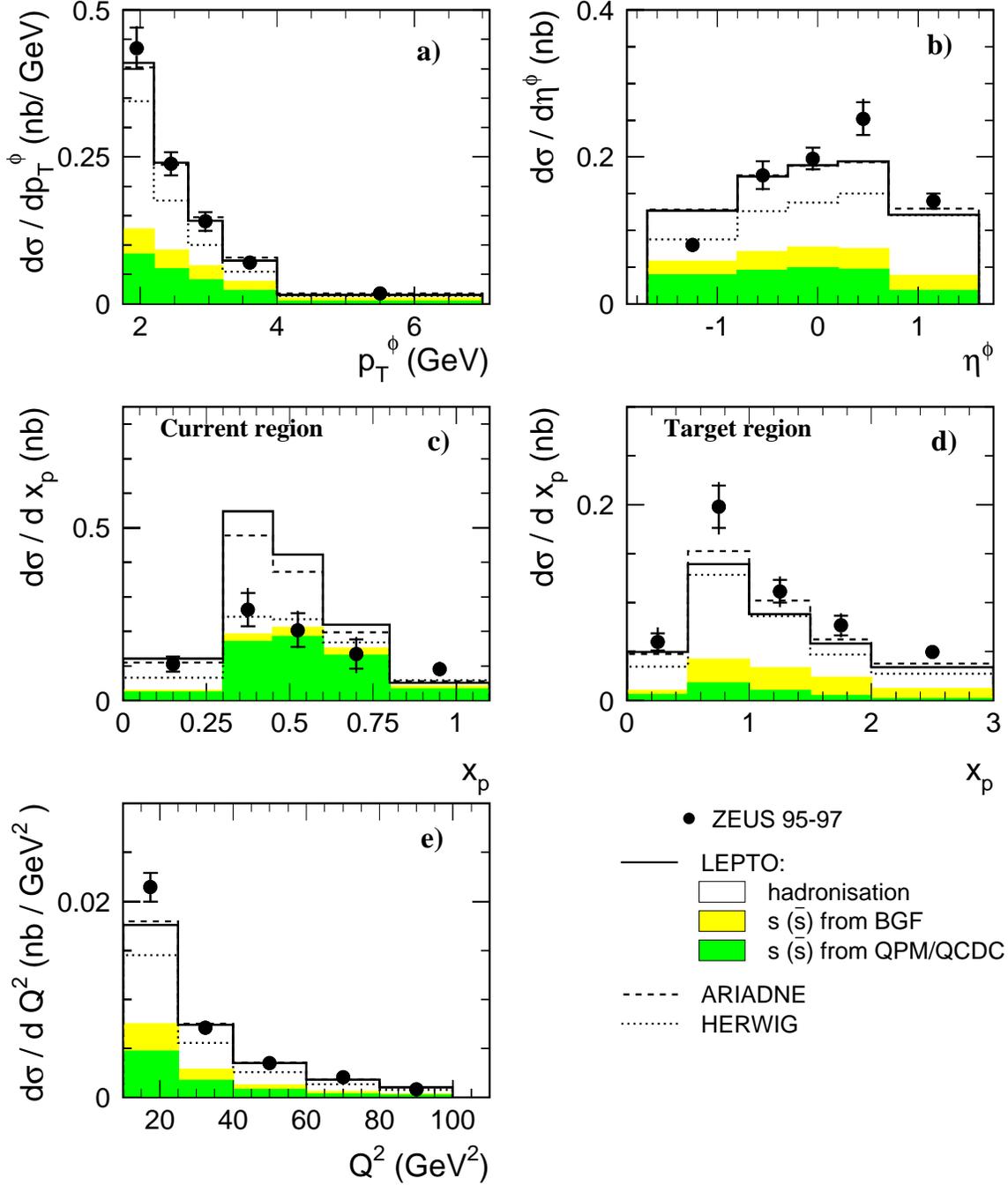}%
\caption{
Differential $\phi$-meson cross sections
as functions of a) $p_T^{\phi}$, b) $\eta^{\phi}$, c)-d) $x_p$ and e) $Q^2$,
compared to LEPTO, ARIADNE and HERWIG. 
The LEPTO and ARIADNE  predictions are shown for $\lambda_s=0.22$.
The data are also compared to contributions from LEPTO events 
with $\phi$ mesons produced
in hard interactions ($s$ or $\bar{s}$ from BGF (light shaded area),
from QPM/QCDC (dark shaded area))
and from events without strange quarks at the parton level
(unshaded area). 
The full error bars include the systematic
uncertainties, which are typically negligible compared to the
statistical errors.}
\label{fig3}
\end{center}
\end{figure}

Figure~\ref{fig3} shows the differential 
cross sections as a function of $p_T^{\phi}$, $\eta^{\phi}$, 
$x_p$ and
$Q^2$ compared to the LEPTO, ARIADNE  and HERWIG models using the 
CTEQ5D parton distribution 
functions\footnote{The 
leading-order set, CTEQ5L, 
gave very similar results.}. 
The measured cross sections are compiled in Table~\ref{table_all}a) - e).   
The $x_p$ 
cross sections are shown  separately for  the current
and the target regions.
The $\phi$-meson cross sections in the current and the target regions 
of the Breit frame are distinctly different:
the data are concentrated at $x_p$ around $\sim 0.5$ 
in the current region, and at $\sim 1$ in the target region.  

The MC models based on the Lund 
fragmentation with $\lambda_s=0.22$ reasonably well reproduce
the $p_T^{\phi}$ and  $Q^2$ distributions.
Significant differences exist 
for the distributions of $\eta^{\phi}$ in the laboratory frame 
and $x_p$  
in the current region of the Breit frame. 
In the target region, the MC models underestimate   
the cross sections. If $\lambda_s=0.3$ is
used, the MC models based on the string fragmentation 
agree well with the data in the target region, but significantly
overestimate the cross sections in the current region.

In addition to varying the $\lambda_s$ values,
different methods to tune the Lund MC models were considered,
all of which had  a negligible effect on the LEPTO
predictions. In particular, 
the contribution to the $\phi$ cross section
from charm events, mainly due to $D^±$ and $D_s^±$ decays,
was investigated using  AROMA \cite{cpc:101:135}. 
This model  produces  charm quarks  exclusively 
through the BGF mechanism, and     
reproduces the measured $D^{*\pm }$ cross sections 
in DIS \cite{zfp:c72:593,*epj:c12:35,*pl:b528:199}.  
According to AROMA, charm decays 
account for $20\%$ of the $\phi$ mesons, contributing 
mainly in the target hemisphere. This fraction is larger than that predicted
by LEPTO, but it is not sufficient to explain the 
observed discrepancies.        
For leading $\phi$ mesons ($\xf>0.8$) in the current region, charm events
give a negligible contribution.

In order to disentangle the different contributions to the
$\phi$-meson production and to investigate           
the observed discrepancies,
the MC samples were divided into a few subsamples.
Figure~\ref{fig3} illustrates 
the contributions of QPM/QCDC interactions 
on an  $s$ or $\bar{s}$ quark of the proton sea.
In this case,  a struck $s$ or $\bar{s}$   
quark produces a $\phi$ meson after the hadronisation
process. 
The $\phi$-meson cross section in the current region of the
Breit frame contains a significant fraction of events produced
by
 hard scatterings of the virtual photon on the strange sea.  
This fraction rises with increasing $p_T^{\phi}$ 
and $x_p$ values, while the  
contribution to $\phi$-meson production from
strange quarks  produced solely in the hadronisation process
becomes negligible for $x_p > 0.8$.  
In contrast, the target region contains a 
small contribution from the QPM/QCDC
events, since the second $s$ or $\bar{s}$  from 
an $s\bar{s}$ pair participating in the interaction 
usually escapes undetected  
in the very forward region.

Figure~\ref{fig3} also 
indicates the contribution of BGF processes  
in which the flavour of the produced quark 
is $s$ or $\bar{s}$.     
The fraction of these BGF events is larger in the target region 
than in the current region.    

\subsection{The $\phi$-meson cross  section  as a function of $x$ }

Production of $\phi$ mesons was investigated 
as a function of  $x$.  
The $s$-quark density increases with decreasing $x$; however, 
the BGF contribution also increases with 
decreasing $x$ due to the rise of the gluon density.   
Thus, the $\phi$-meson cross section 
as a function of $x$ depends on both 
the strange sea and the gluon density.

\begin{figure}
\begin{center}
\includegraphics[height=18.0cm]{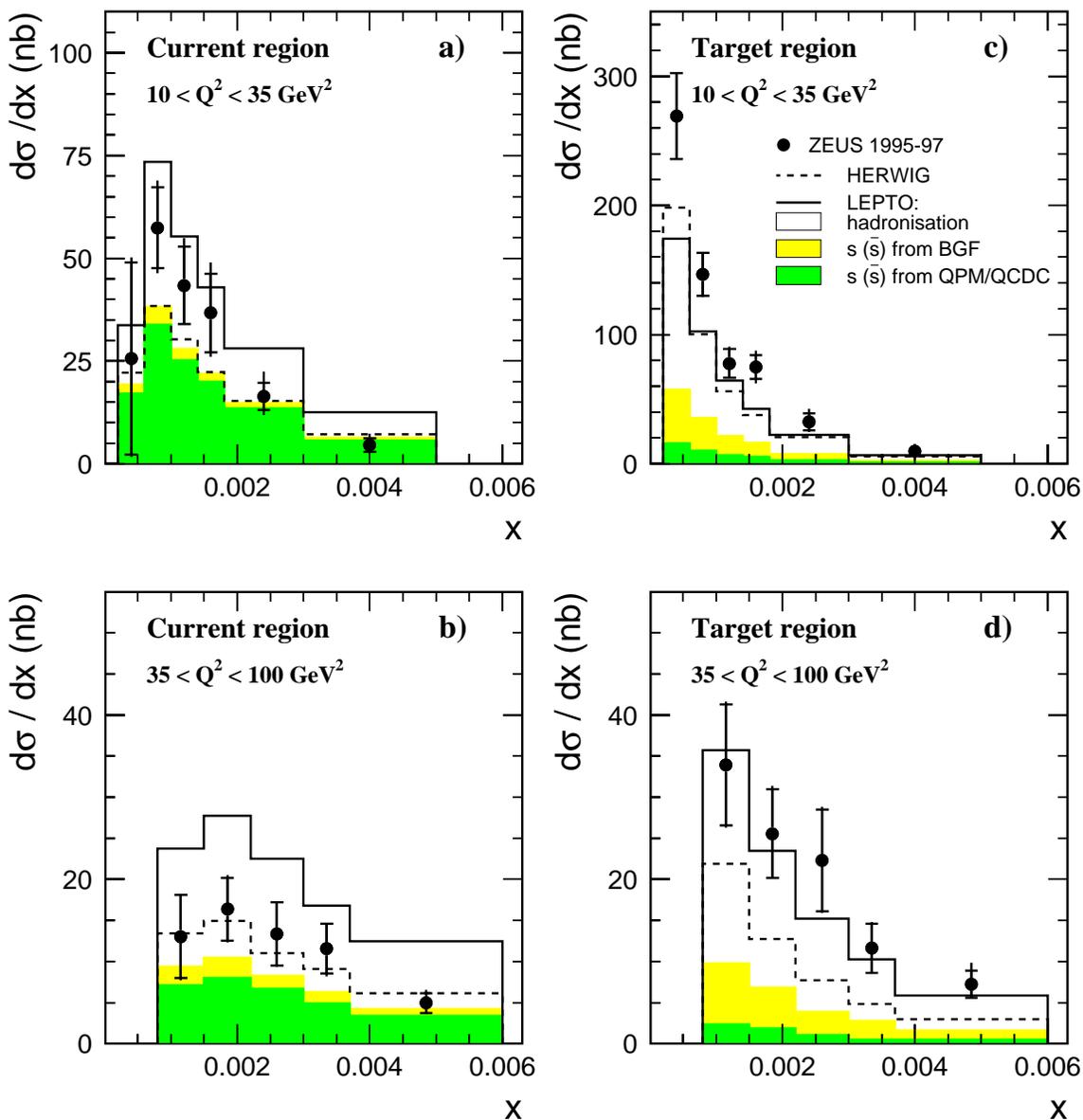}
\caption{
The inclusive sections as a function of $x$ for two $Q^2$ intervals,
for the  current, a)-b), 
and  the
target, c)-d),  regions of the Breit frame
compared to the HERWIG (dashed lines) and the LEPTO (solid lines) predictions
with $\lambda_s=0.22$.
The LEPTO model shows
separately the contributions from events with $\phi$ mesons produced
in hard interactions ($s$ or $\bar{s}$ from BGF (light shaded area), 
from QPM/QCDC (dark shaded area))
and from events without strange quarks at the parton level
(unshaded area). }
\label{fig4}
\end{center}
\end{figure}

The differential cross sections
as a function of $x$   
for two $Q^2$ regions, $10<Q^2<35$~GeV$^2$ and 
$35<Q^2<100$~GeV$^2$, are shown in Fig.~\ref{fig4}.
Table~\ref{table_f1} 
gives the values of the cross sections.
The $\phi$-meson differential cross section increases  
as $x$ decreases down to the kinematic limit. 
The LEPTO and HERWIG MCs reproduce this rise.  
The LEPTO model shows 
the contributions of events
in  which  a $\phi$ meson is produced after hadronisation
of an $s$ ($\bar{s}$) quark emerging from the hard interaction.
The contributions from the QPM/QCDC and BGF
processes rise with decreasing $x$
due to the  rise
of the $s$-quark and the gluon density in the proton.

\subsection{Leading $\phi$ mesons}

The MC predictions for leading $\phi$ mesons ($\xf > 0.8$),
usually 
corresponding to high $p_T^{\phi}$ in the laboratory frame,  
have small uncertainties both  
in the simulation
of the QCD processes and in the hadronisation mechanism;
for a given strange-sea density,
the scattering of the virtual photon on a strange 
quark  is described by the QED process, $\gamma^*s\to s$.
Any additional gluon emissions are not important for $\xf > 0.8$,
since such processes lead to strange quarks with smaller $\xf$.

\begin{figure}
\begin{center}
\includegraphics[height=14.0cm]{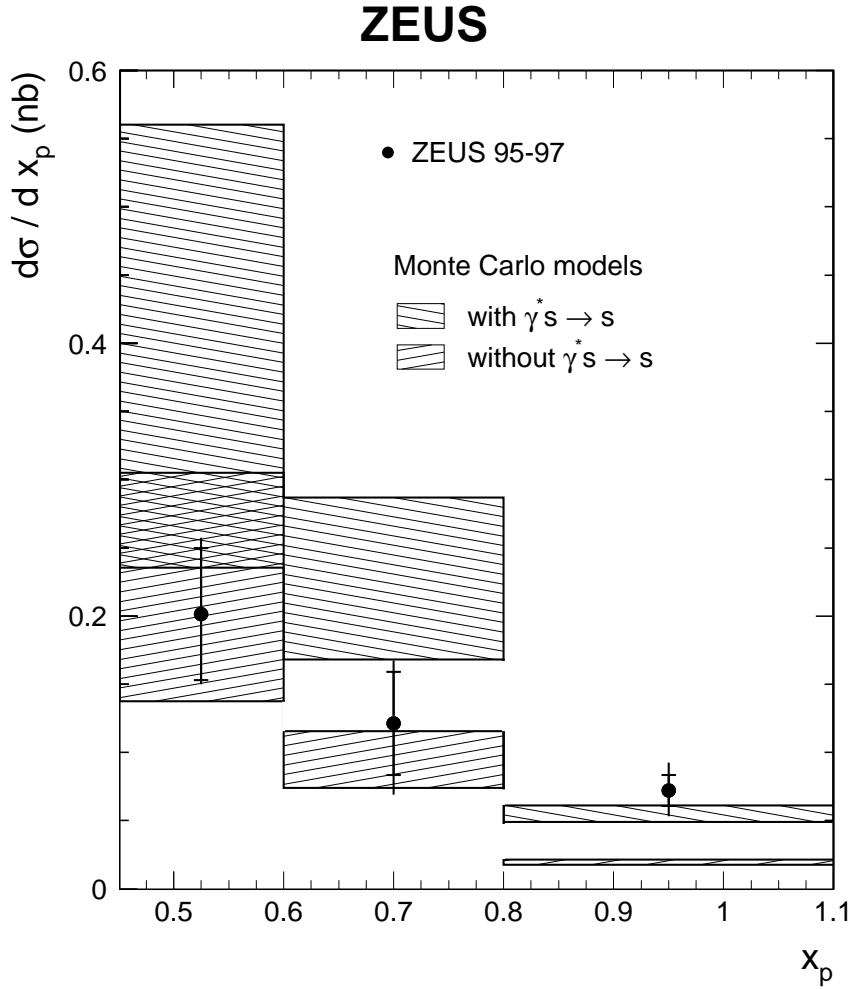}%
\caption{
The cross sections for leading  $\phi$ mesons 
as a function of $\xf$ 
in the current
region of the Breit frame for  
$\eta_{\mathrm{max}}>2$.  
The hatched bands represent uncertainties in the simulation of
the $\phi$-meson production by Monte Carlo models,
and include LEPTO ($\lambda_s=0.2-0.3$), ARIADNE and HERWIG.
The upper bounds
of the hatched area  
correspond to  LEPTO with $\lambda_s=0.3$,
while the lower  bounds of this area are defined by 
the LEPTO ($\lambda_s=0.2$) and HERWIG predictions (see text). 
}
\label{fig5}
\end{center}
\end{figure}

Figure~\ref{fig5} and Table~\ref{table_all}f) show the 
cross sections for three $x_p$ bins in  
the current region of the Breit frame for the full $Q^2$ range, 
after removing the diffractive contribution with $\eta_{\mathrm{max}}<2$.  
The hatched bands represent uncertainties in the simulation of
the $\phi$-meson production by the MC models
LEPTO, ARIADNE and HERWIG. The uncertainty  due to
$\lambda_s$ values between 0.2 and 0.3 
is also included, such that the upper bounds
of the hatched area for $\xf<0.8$ correspond to  LEPTO with $\lambda_s=0.3$,
while the lower  bounds of this area indicate 
the HERWIG predictions. For $\xf >0.8$, the HERWIG prediction is between 
LEPTO with $\lambda_s=0.2$ and  $\lambda_s=0.3$. 
The predicted cross sections of  ARIADNE
are always within the shaded bands.

The MC uncertainties are small for $\xf>0.8$. 
The predictions are shown with and without the contribution
from the process of Fig.~\ref{fey}a).
The measured cross section clearly 
requires a contribution from
interactions with the strange sea.
The MCs with the CTEQ5D or the MRST99(c-g)\cite{epj:c4:463} (not shown) 
parton distribution functions reproduce the measured
rate of $\phi$ mesons. In these parameterisations, the strange sea 
is suppressed with respect to the non-strange sea, with      
the ratio $s\bar{s}/d\bar{d}$ in the range $0.25-0.5$, depending on $x$. 
The predictions correctly describe the results and thus confirm
the strange-quark suppression, even though 
the $Q^2$ values of this data are significantly larger
than the strange-quark mass.  

\section{Conclusions}

Inclusive $\phi$-meson cross sections  
have been measured in deep inelastic scattering
for $10<Q^2<100$~GeV$^2$,
$2\cdot 10^{-4} < x < 10^{-2}$,
$1.7 < p_T^{\phi} <7$~GeV and $-1.7 < \eta^{\phi } < 1.6$. 
The MC predictions with 
a strangeness-suppression factor
$\lambda_s=0.3$ overestimate
the measured cross sections.  
A smaller value of the strangeness-suppression factor, 
$\lambda_s=0.22\pm 0.02$,    
reduces the predicted cross sections and gives a good   
description of the total $\phi$-meson cross section, as well as   
of the differential   
$p_T^{\phi}$, $Q^2$ and $x$ cross sections.
However, Monte Carlo models  based on Lund fragmentation 
fail to describe 
the $\eta^{\phi}$ and the 
$x_p$ cross sections.
The HERWIG simulation describes 
the measured cross section in the current region well, 
but predicts a smaller 
overall cross section than that measured;   
$\phi$-meson production in the target region is underestimated 
by all MC models. 

The production of $\phi$ mesons in the current region
of the Breit frame has a significant contribution  from
the hard scattering
of a virtual photon on the strange sea of the proton.
The predictions for the rate of high-momentum $\phi$ mesons with large values of
the scaled momentum, $x_p>0.8$, in the current region of the Breit frame
have small uncertainties, since the $\phi$ production in this region 
is dominated by $\gamma^*s \to s$ scattering. 
To reproduce the observed rate of $\phi$ mesons at high $\xf$, the MC
models require a significant contribution from the
strange sea of the proton.
In this region, the measured cross section is correctly reproduced 
by these models when $\gamma^*s \to s$ scattering is included.
These results constitute the first direct evidence for
the existence of the strange sea in the proton at  $x<0.006$.

\section*{Acknowledgments}
\vspace{0.3cm}
We thank the DESY Directorate for their strong support and encouragement.
The remarkable achievements of the HERA machine group were essential for
the successful completion of this work and are greatly appreciated. We
are grateful for the support of the DESY computing and network services.
The design, construction and installation of the ZEUS detector have been
made possible owing to the ingenuity and effort of many people from DESY
and home institutes who are not listed as authors.

\vfill\eject

\providecommand{\etal}{et al.\xspace}
\providecommand{\coll}{Collaboration}


\vspace{2.0cm}

\begin{table}[p]
\begin{center}
\begin{tabular}{|r|l|r|l|}
\hline
a) \hfill range (GeV)& $d\sigma / dp_T^{\phi}$ (nb/GeV) &  b)  \hfill range & $d\sigma / d \eta^{\phi}$ (nb) \\ \hline 
$ 1.7- 2.2$&$0.433\pm 0.035^{+0.014}_{-0.001  }$ & $-1.70--0.80$&$0.079\pm 0.009^{+0.002}_{-0.004}$ \\ 
$ 2.2- 2.7$&$0.237\pm 0.019^{+0.002}_{-0.007  }$ & $-0.80--0.30$&$0.175\pm 0.019^{+0.002}_{-0.007}$ \\  
$ 2.7- 3.2$&$0.142\pm 0.016^{+0.002}_{-0.010  }$ & $-0.30- 0.20$&$0.199\pm 0.014^{+0.003}_{-0.007}$ \\  
$ 3.2- 4.0$&$0.071\pm 0.009^{+0.004}_{-0.001  }$ & $ 0.20- 0.70$&$0.248\pm 0.022^{+0.020}_{-0.001}$ \\  
$ 4.0- 7.0$&$0.018\pm 0.002^{+0.001}_{-0.002  }$ & $ 0.70- 1.60$&$0.140\pm 0.010^{+0.001}_{-0.003}$ \\  
\hline\hline
c) \hfill range (GeV$^2$)& $d\sigma /dQ^2$ (nb/GeV$^2$) & d)  \hfill range & $d\sigma / d\xf (\mathrm{target})$ (nb)   \\ \hline 
$ 10- 25$&$0.02140\pm 0.00148^{+0.00024}_{-0.00031}$ & $0.0- 0.5$&$0.059\pm 0.009^{+0.013}_{-0.007 }$\\ 
$ 25- 40$&$0.00708\pm 0.00047^{+0.00019}_{-0.00002}$ & $0.5- 1.0$&$0.193\pm 0.021^{+0.031}_{-0.034 }$ \\ 
$ 40- 60$&$0.00350\pm 0.00040^{+0.00008}_{-0.00001}$ & $1.0- 1.5$&$0.112\pm 0.012^{+0.006}_{-0.012 }$ \\ 
$ 60- 80$&$0.00199\pm 0.00026^{+0.00021}_{-0.00004}$ & $1.5- 2.0$&$0.077\pm 0.010^{+0.008}_{-0.007 }$ \\ 
$ 80- 100$&$0.00079\pm 0.00021^{+0.00001}_{-0.00008}$& $2.0- 3.0$&$0.049\pm 0.006^{+0.004}_{-0.004 }$ \\ 
\hline\hline
e)  \hfill range & $d\sigma / d\xf (\mathrm{current})$ (nb) & f) \hfill range($\eta_{\mathrm{max}}>2$)  & $d\sigma / d\xf (\mathrm{current})$ (nb)\\ \hline  
$ 0.00- 0.30$&$0.105\pm 0.022^{+0.008}_{-0.001}$ & $ 0.00- 0.30$&$0.105\pm 0.022^{+0.008}_{-0.001}$ \\
$ 0.30- 0.45$&$0.263\pm 0.048^{+0.040}_{-0.006}$ & $ 0.30- 0.45$&$0.262\pm 0.048^{+0.040}_{-0.006}$ \\
$ 0.45- 0.60$&$0.203\pm 0.049^{+0.028}_{-0.018}$ & $ 0.45- 0.60$&$0.201\pm 0.048^{+0.028}_{-0.018}$ \\
$ 0.60- 0.80$&$0.135\pm 0.042^{+0.026}_{-0.036}$ & $ 0.60- 0.80$&$0.121\pm 0.037^{+0.026}_{-0.035}$ \\  
$ 0.80- 1.10$&$0.090\pm 0.014^{+0.017}_{-0.015}$ & $ 0.80- 1.10$&$0.072\pm 0.011^{+0.016}_{-0.015}$ \\
\hline 

\end{tabular}

\caption{
Differential $\phi$-meson cross sections
as functions of $p_T^{\phi }$, $\eta ^{\phi}$,
$Q^2$ and $\xf$.
The statistical and asymmetric systematic uncertainties 
are shown separately.
}
\label{table_all}
\end{center}
\end{table}

\newpage

\begin{table}  
\begin{minipage}[c]{0.5\textwidth}%

\begin{center}
$10<Q^2<35$ GeV$^2$
\begin{tabular}{|l|l|}
\hline
range (all regions) & $d\sigma / dx  $ (nb)\\ \hline\hline
$0.0002-0.0006$&$288.6\pm  33.0^{+  7.5}_{- 20.8  }$ \\ \hline
$0.0006-0.0010$&$216.7\pm  24.8^{+  4.6}_{-  0.5  }$ \\ \hline
$0.0010-0.0014$&$110.0\pm   8.5^{+  5.6}_{-  3.0  }$ \\ \hline
$0.0014-0.0018$&$105.7\pm  17.3^{+  1.9}_{-  4.9  }$ \\ \hline
$0.0018-0.0030$&$ 47.8\pm   5.3^{+  0.8}_{-  3.1  }$ \\ \hline
$0.0030-0.0050$&$ 11.7\pm   3.1^{+  0.9}_{-  0.1  }$ \\ \hline
Current region \\  
\hline
 range & $d\sigma / dx $ (nb)\\ \hline\hline
$0.0002-0.0006$&$ 25.6\pm  23.4^{+  7.6}_{-  4.7  }$ \\ \hline
$0.0006-0.0010$&$ 57.4\pm   9.9^{+  5.9}_{-  5.2  }$ \\ \hline
$0.0010-0.0014$&$ 43.4\pm   9.4^{+  6.6}_{-  0.3  }$ \\ \hline
$0.0014-0.0018$&$ 36.7\pm   9.5^{+  7.7}_{-  4.7  }$ \\ \hline
$0.0018-0.0030$&$ 16.4\pm   3.3^{+  5.0}_{-  3.0  }$ \\ \hline
$0.0030-0.0050$&$  4.5\pm   1.7^{+  2.2}_{-  0.7  }$ \\ \hline
Target region \\ 
\hline
 range & $d\sigma / dx  $ (nb)\\ \hline\hline
$0.0002-0.0006$&$269.0\pm  33.4^{+  2.8}_{-  4.7  }$ \\ \hline
$0.0006-0.0010$&$146.1\pm  16.6^{+  0.5}_{-  4.7  }$ \\ \hline
$0.0010-0.0014$&$ 78.5\pm  11.3^{+  6.9}_{-  0.7  }$ \\ \hline
$0.0014-0.0018$&$ 74.7\pm   9.2^{+  8.9}_{-  9.2  }$ \\ \hline
$0.0018-0.0030$&$ 32.4\pm   6.6^{+  7.5}_{-  4.9  }$ \\ \hline
$0.0030-0.0050$&$  9.4\pm   5.2^{+  4.8}_{-  2.6  }$ \\ \hline
 \end{tabular}

\end{center}

\end{minipage}%
\begin{minipage}[c]{0.5\textwidth}%

\begin{center}
$35<Q^2<100$ GeV$^2$
 \begin{tabular}{|l|l|}
 \hline
 range (all regions) & $d\sigma / dx $ (nb)\\ \hline\hline
$0.0008-0.0015$&$ 45.9\pm  13.0^{+  0.5}_{-  2.9  }$ \\ \hline
$0.0015-0.0022$&$ 39.0\pm   4.6^{+  1.7}_{-  0.1  }$ \\ \hline
$0.0022-0.0030$&$ 36.0\pm   8.2^{+  0.8}_{-  0.1  }$ \\ \hline
$0.0030-0.0037$&$ 22.0\pm   3.4^{+  0.4}_{-  5.6  }$ \\ \hline
$0.0037-0.0060$&$ 10.0\pm   1.8^{+  0.2}_{-  0.1  }$ \\ \hline
Current region \\ 
\hline
  range & $d\sigma / dx $ (nb)\\ \hline\hline
$0.0008-0.0015$&$ 13.0\pm   5.1^{+  1.0}_{-  1.6  }$ \\ \hline
$0.0015-0.0022$&$ 16.4\pm   3.8^{+  1.6}_{-  1.3  }$ \\ \hline
$0.0022-0.0030$&$ 13.4\pm   3.8^{+  0.9}_{-  1.1  }$ \\ \hline
$0.0030-0.0037$&$ 11.5\pm   3.0^{+  0.6}_{-  1.4  }$ \\ \hline
$0.0037-0.0060$&$  4.9\pm   1.2^{+  1.0}_{-  0.7  }$ \\ \hline
Target region \\ 
\hline
  range & $d\sigma / dx $ (nb)\\ \hline\hline
$0.0008-0.0015$&$ 34.3\pm   7.5^{+  2.2}_{-  2.1  }$ \\ \hline
$0.0015-0.0022$&$ 25.7\pm   5.4^{+  1.2}_{-  1.2  }$ \\ \hline
$0.0022-0.0030$&$ 22.2\pm   6.2^{+  2.1}_{-  1.4  }$ \\ \hline
$0.0030-0.0037$&$ 11.7\pm   3.0^{+  0.9}_{-  0.7  }$ \\ \hline
$0.0037-0.0060$&$  7.2\pm   1.7^{+  2.0}_{-  0.1  }$ \\ \hline
 \end{tabular}

\end{center}

\end{minipage}%
\caption{
Differential $\phi$-meson cross sections
as a function of $x$ for two intervals in $Q^2$.
The statistical and asymmetric systematic uncertainties  
are shown separately.
}
\label{table_f1}
\end{table}


\newpage

\newpage 

\newpage

%
%
\end{document}